\UseRawInputEncoding 
\documentclass[%
twocolumn,superscriptaddress,
 amsmath,amssymb,
 aps,
prl,
]{revtex4}
\usepackage{mathrsfs,amsmath,amssymb,amsthm}
\usepackage{amssymb,fmtcount}
\usepackage{graphicx,epsfig,latexsym,overpic,amssymb,color}
\usepackage{threeparttable}
\preprint{\today}

\begin{document}
\title{
Fission Dynamics of Compound Nuclei: Pairing versus Fluctuations
}
\author{Yu Qiang}
\affiliation{
State Key Laboratory of Nuclear Physics and Technology, School of Physics,
Peking University, Beijing 100871, China
}
\author{J.C. Pei}\email{peij@pku.edu.cn}
\affiliation{
State Key Laboratory of Nuclear Physics and Technology, School of Physics,
Peking University, Beijing 100871, China
}
\author{P.D. Stevenson}
\affiliation{
Department of Physics, University of Surrey, Guildford, Surrey, GU2 7XH, United Kingdom
}

\begin{abstract}

Energy dependence of fission observables is a key issue for wide nuclear applications.
We studied real-time fission dynamics from low-energy to high excitations in the compound nucleus $^{240}$Pu
with the time-dependent Hartree-Fock+BCS approach. It is shown that the evolution time of
the later phase of fission towards scission is considerably lengthened at finite temperature.
As the role of dynamical pairing is vanishing at high excitations, the random transition between single-particle levels around the Fermi surface to mimic thermal fluctuations is indispensable
to drive fission. The obtained fission yields and total kinetic energies with fluctuations can be divided into two asymmetric scission channels, namely S1 and S2, which explain well experimental results, and give microscopic support to the Brosa model. With increasing fluctuations, S2 channel takes over S1 channel and the spreading fission observables are obtained.

\end{abstract}
\maketitle

A deeper understanding of fission from a droplet of condensed nuclear matter splitting into fragments is still strongly motivated,
even though its discovery occurred more than 80 years ago~\cite{meitner}.
Firstly,  fission studies are crucial for increasingly wide nuclear applications~\cite{wp},
as well as for basic sciences such as synthesis of superheavy elements~\cite{she,pei2009}
and constraints on $r$-process in neutron-star mergers~\cite{rp1,rp2,witek}.
However, fission measurements are very difficult and energy dependent fission data is sparse
in major nuclear data libraries~\cite{endf}.
Secondly, the fission process is extremely complex from the microscopic view as a probe of non-equilibrium quantum many-body dynamics~\cite{yorkreview,Schmidt,schunck}.

It is known that the pioneer Bohr-Wheeler statistical theory is very successful but not applicable for highly excited fission
with experimental observations of exceeding prescission neutron multiplicities~\cite{bertsch}.  Strong viscosity and dissipation in hot nuclear matter has to be invoked~\cite{dipole}.
The realistic fission of compound nuclei is not only determined by the barrier but also the later phase
of fission evolutions towards scission becomes important~\cite{fluc-diss}. In addition,  the quantum effects such as shell effects and pairing would gradually fade away
as excitation energies increase~\cite{pei2009,zhuy}.
There were studies based on temperature-dependent fission barriers~\cite{pei2009,zhuy,tdgcm2} or energy-dependent level densities~\cite{2017}, however, a fully microscopic fission dynamics
 in terms of excitation energy dependence is still absent.

For experiments, the well-known semi-empirical model by Brosa {\it{et al.}}~\cite{brosa}
is the  primary tool for evaluations of fission data with high accuracy.
This model  has great physics intuition on the multi-channel fission and the random neck-rupture assumptions,
which is well established by detailed fission observations, in particular
correlations between distributions of mass yields, total kinetic energies (TKE) and neutron multiplicities.
However, as a major obstacle for extrapolations when experiments are absent, the origin and pathways of two
asymmetric standard channels (denoted S1, S2) in Brosa model are ambiguous, although shell effects are present in nascent fragments~\cite{Schmidt,scamps}.
For shape dynamics models~\cite{tdgcm,tdgcm2,langevin,langevin2,langevin3}
based on complex potential energy surfaces (PES), it is still difficult to identify
pathways of these two modes.
Therefore, the validation of physics assumptions of the Brosa model from microscopic dynamical models would be significant.

The microscopic time-dependent density functional theory (TD-DFT) is promising to describe the later phase of fission from saddle to scission~\cite{koonin,negele,tddft,tdhf3,bulgac2016}.
TD-DFT has provided valuable clues about the overdamped assumption~\cite{bulgac2}, non-adiabatic effects~\cite{258fm},  the excitations of fragments~\cite{bulgac2016}, the role of shell effects~\cite{scamps} and pairing effects~\cite{bulgac2016},
but the lack of fluctuations undermines TD-DFT to reproduce distributions of fission yields~\cite{lacroix2014,bulgac2}.
It is an evident defect that strongly dissipated fission has no dissipation-fluctuation correspondence.
At low excitations, the probability of orbital exchanges is connected to the Landau-Zener effect and is dependent on the pairing gap~\cite{koonin}.
At high excitations, as the pairing is vanishing,  it is expected that thermal fluctuations are the main source of orbital changes.
There are efforts such as the stochastic TD-DFT with initial fluctuations~\cite{stdhf,lacroix2014}  or by including dynamical
density fluctuations~\cite{bulgac3}, aiming to bridge the Langevin descriptions~\cite{langevin}.
The time-dependent random-phase approximation~\cite{tdrpa} can describe particle-number fluctuations
but not actual distributions of fission observables~\cite{176Yb,258fm}.
In addition to quantal fluctuations,
it is essential to include thermal fluctuations based on TD-DFT which would become significant in fission of compound nuclei.
Thermal fluctuations in the mean-field picture can be naturally linked to
random transitions between single-particle levels around Fermi surfaces.
Actually the fluctuations in single-particle and collective motions
are interweaved in the TD-DFT approach.

In this Letter, we study the energy dependence of various fission observables of the compound nucleus  $^{240}$Pu with mircoscopic TD-DFT, including dynamical pairing
and thermal fluctuations. This is an attempt to develop a
unifying fission framework by connecting microscopic dynamical models and statistical Langevin models. As a reward,  it turns out that our results can
explain the origin of the two asymmetric fission channels of the Brosa model.

We describe the fission of compound nuclei with the time-dependent Hartree-Fock+BCS (TD-BCS) approach~\cite{TDHFB,TDBCS}.
The initial configuration of compound nuclei $^{240}$Pu is obtained by finite-temperature Hartree-Fock+BCS calculations~\cite{FT-HFB,zhuy}.
The evolutions of compound nuclei is similar to that of the zero-temperature time-dependent Hartree-Fock-Bogoliubov (TD-HFB) formulism~\cite{TDHFB},
\begin{equation}
i\hbar \frac{\textrm{d} \mathcal{R}}{\textrm{d} t}= [H, \mathcal{R}],
\end{equation}
where $H$ is the HFB hamiltonian,  $\mathcal{R}$ is the general density matrix.
The initial $H$ and $\mathcal{R}$ are associated with a finite temperature~\cite{FT-HFB}.
The time-dependent Hartree-Fock+BCS equations can be obtained by using BCS basis
or canonical basis~\cite{TDHFB,TDBCS}.
Note that TD-BCS can describe dynamical pairing approximately compared to the fully dynamical pairing in
TD-HFB.

In TD-BCS, the evolution of densities is actually related to the evolution of occupation numbers of single-particle levels. In the mean-field picture,
the single-particle levels around Fermi surfaces are active for orbital exchanges due to dynamical pairing fluctuations~\cite{bulgac2016}.
To mimic thermal fluctuations, we implement random transitions between single-particle levels without explicit external forces,
in which the occupation number $n_k$ is modified with a random additive $\delta n_k$.
The random $\delta n_k$ is designed as a transition so that the total particle number is  strictly conserved.
The transition occurs as a random Gaussian noise around Fermi surfaces.
The transition amplitude $\delta n_{kj} = q_r C_{kj}\exp(-\frac{|\varepsilon_k-\varepsilon_j|}{T})$,
where $e^{(-|\varepsilon_k-\varepsilon_j|/T)}$ is a symmetric Boltzmann distribution, $T$ is an effective temperature and $\varepsilon_{k,j}$ are single-particle energies.
The transition amplitudes are also constrained by the Pauli exclusion principle. For two levels with occupation numbers $n_k$ and $n_j$, we take $C_{kj}$=min$(n_k, 1-n_j)$ which is the maximum allowed symmetric transition amplitude.
The transition occurs randomly as jump up or down determined by another random number $q_r \in [-1.0, 1.0]$.
The random transitions simulate nucleon-nucleon collisions and also play as a remedy to truncated correlations, while the exact treatment of collision terms beyond TDHF is
very sophisticated~\cite{etdhf}. The effective temperature in fluctuations is not necessary the initial temperature of compound nuclei.
Note that fluctuations can be considerable even in spontaneous fission~\cite{2016}.
At high temperatures,   the orbital exchanges are mainly induced by thermal fluctuations,
 even when two levels are not close.

The calculations are performed with the time-dependent Hartree-Fock solver Sky3D~\cite{sky3d,sky3d1} with the addition of our
modifications of TD-BCS plus thermal fluctuations. The initial configurations at finite temperatures are obtained using the SkyAx solver~\cite{skyax,zhuy}, to interface with Sky3D~\cite{marko}.
The excitation energy of compound nuclei is related to the initial temperature~\cite{FT-HFB,pei2009}.
 The time evolution operator is based on the Taylor expansion at the fourth order
and time step is taken as 0.1 fm/c. The box size (x, y, z) is taken as 48$\times$48$\times$64 fm and the grid space is 0.8 fm.
The nuclear interaction we adopted is the widely used SkM$^{*}$ parameterization~\cite{skm} and the paring interaction is the mixed pairing~\cite{mix-pair}.
More details of the methods are given in the supplement~\cite{supp}.

\begin{figure}[t]
\centering
\includegraphics[width=0.47\textwidth]{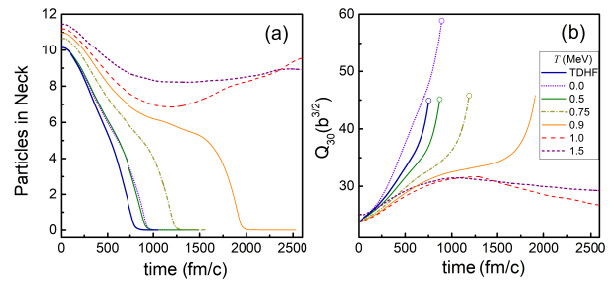}
\caption{
The TD-BCS evolutions of the number of particles in a neck of 2 fm length at the density minimum (a) and  the octupole deformation $Q_{30}$ (in units of $b$=100 fm$^2$) (b) of  $^{240}$Pu with different initial temperatures $T$.
 TDHF results without pairing are also shown.                                                   \label{FIG1}
}
\end{figure}

\begin{figure}[b]
\centering
\includegraphics[width=0.45\textwidth]{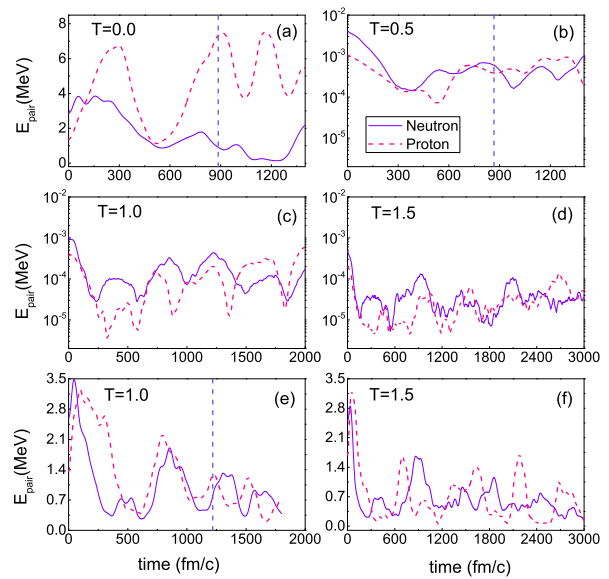}
\caption{
The evolutions of neutron and proton pairing energies (in MeV) within TD-BCS at different initial temperatures of $T$=0 MeV (a), 0.5 (b), 1.0 (c), 1.5 (d).
TD-BCS results with the initial pairing field at zero temperature are also shown for $T$=1.0 MeV (e), 1.5 (f).                                                      \label{FIG2}
}
\end{figure}

We firstly studied the fission of compound nuclei $^{240}$Pu with different initial temperatures with TD-BCS.
The initial deformation in this work adopts the dimensionless quadrupole-octupole deformations as $\beta_2$=2.3 and $\beta_3$=1.0 (see the definition~\cite{bender}).
The timescale is an important quantity characterizing nuclear dynamics with dissipations and fluctuations~\cite{timescale}.
Fig.~\ref{FIG1}(a) displays the evolutions of the number of particles in the neck.
The zero-temperature TD-BCS calculations is slower than  TDHF calculations
 due to its longer fission pathway.
With increasing temperatures $T$, the evolution times
become considerably lengthened.
Note that fission would not occur above $T$=0.9 MeV within TD-BCS.
At $T$=0.9 MeV, corresponding to an excitation energy of 16.1 MeV,  the evolution takes 1900 fm/c, or 6.3$\times$10$^{-21}$s.
It can be seen that timescales of the later phase of fission are indeed considerable compared to the statistical model at high excitations.
 For example, the timescale is about 10$^{-20\sim21}$ s by statistical models for fission of highly excited superheavy nuclei~\cite{feng}.
Fig.~\ref{FIG1}(b) displays the evolutions of  octupole deformations.  TD-BCS calculations at zero temperature result in a larger octupole deformation at scission.
At high excitations, shape evolutions become very slow as an indication of increasing viscosities.

\begin{figure}[t]
\centering
\includegraphics[width=0.45\textwidth]{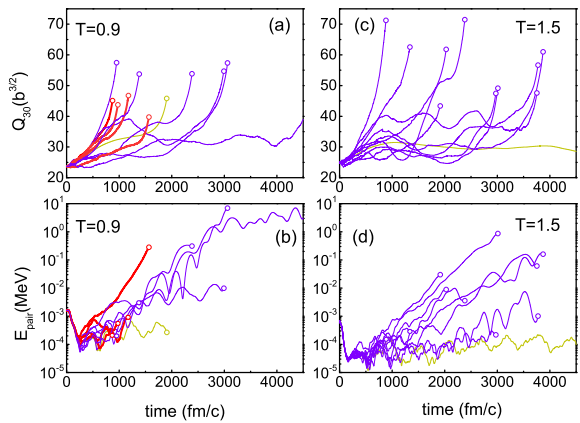}
\caption{
The evolutions of  $Q_{30}$ and pairing energies within TD-BCS plus thermal fluctuations at $T$ of 0.9 and 1.5 MeV.
Results of $T$=0.9 with small scission deformations (S1 channel) are shown in thick red lines.
Results without fluctuations in yellow (light-grey) lines are also shown for comparisons.
\label{FIG3}
}
\end{figure}

It is of great concern about the role of dynamical pairing in fission of compound nuclei when the initial static pairing is vanishing.
Fig.~\ref{FIG2} displays the
evolution of pairing energies with different initial temperatures.
We see that both initial pairing and dynamical pairing are very small at high excitations.
Note that the pairing above the critical temperature is not strictly zero~\cite{khan}.
 It was known that in some cases fission
 can happen within TD-BCS or TD-HFB but not within TDHF~\cite{258fm,paul}.
For tests, we also preformed TD-BCS calculations at high temperatures but with an initial pairing field at zero temperature,
see Fig.~\ref{FIG2}(e,f).
Dynamical pairing fluctuations in Fig.~\ref{FIG2}(e,f) are suppressed compared to zero-temperature results in Fig.~\ref{FIG2}(a).
 In hot nuclei, we see that pairing  energies dissipate rapidly at the beginning stage of evolutions.
Furthermore,  the damping time from $T$=1.0 to 1.5 MeV decreases, which indicates increasing viscosities and dissipations as temperature increases.
This also implies that the initial lubricant pairing can reduce viscosity to some extent.
The fission now happens at $T$=1.0 and 1.25 MeV with initial pairings, but still not happen at $T$=1.5.
 In this case, thermal fluctuations have
to be invoked.

\begin{table}[b]
  \caption{ Calculated fission observables of $^{240}$Pu at different initial temperatures $T$ (MeV) and associated excitation energies $E^{*}$, including mass of heavy fragment $A_H$, excitation energies of heavy fragments $E_{H}^{*}$ and light fragments $E_{L}^{*}$, and TKE. All energies are in MeV.
  TDHF results are also listed. TD-BCS results with an initial pairing of zero temperature are listed for comparisons.
  With thermal fluctuations,  averaged values and standard deviations in brackets are shown.}
 \begin{tabular}{lcccccc}
  \hline
  \hline
 \hspace{7pt}  $T$ ($E^{*}$) \hspace{6pt}  & \hspace{6pt}  $A_H$ \hspace{11pt}  & \hspace{11pt}  $E_{H}^{*}$ \hspace{11pt}  & \hspace{11pt} $E_{L}^{*}$ \hspace{11pt}  &  \hspace{11pt}  TKE \hspace{9pt} \\
  \hline
 TDHF & 134.9 & 9.9 & 12.0 & 186.9 \\
   \hline
 \multicolumn{3}{l}{TD-BCS with temperature} \\
  0.5 (4.7) & 135.3 & 9.1 & 19.3 & 186.9 \\
  0.75 (10.6) & 135.8 & 13.8 & 21.3 & 185.3 \\
  0.9 (16.1) & 135.6 & 17.8 & 24.8 &  185.6 \\
 \hline
\multicolumn{3}{l}{with initial pairing} \\
  0.0 & 138.6 & 10.6 & 22.6 & 172.1 \\
  0.75 (10.6) & 137.7 & 13.5 & 25.1  & 175.6 \\
  1.0 (20.5) & 138.4 & 19.8 & 28.8 & 174.2 \\
  1.25 (34.6) & 137.0 & 28.9 & 32.9  & 176.9 \\
   \hline
\multicolumn{3}{l}{with thermal fluctuations} \\
  0.75 (10.6) & 136.5(1.8) & 14.5(2.5) & 24.8(3.2)  & 180.9(6.9) \\
  0.9 (16.1) & 137.5(2.4) & 20.5(3.3) & 27.4(2.7)  & 177.4(6.9) \\
  1.5 (53.2) & 138.5(4.9) & 41.4(5.7) & 42.3(4.9) & 172.6(3.9) \\
  \hline
  \hline
\end{tabular}
  \label{tab1}
\end{table}

\begin{figure}[t]
\centering
\includegraphics[width=0.45\textwidth]{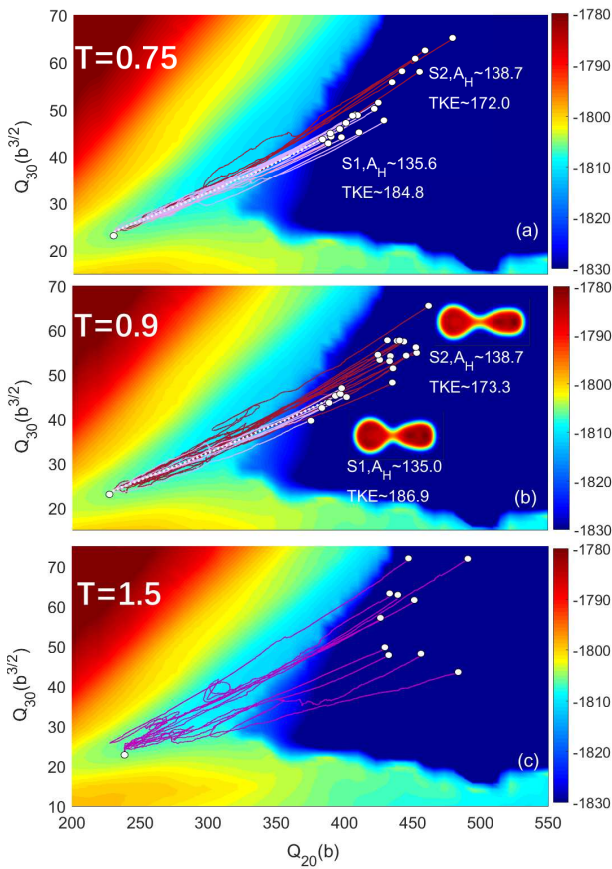}
\caption{
The fission pathways  of $^{240}$Pu within TD-BCS plus thermal fluctuations in the space of quadrupole-octupole deformations ($Q_{20}$, $Q_{30}$),
at temperatures of 0.75 MeV (a), 0.9 MeV (b) and 1.5 MeV (c).
At $T$=0.75 and 0.9 MeV, the fission pathway without fluctuations (dashed white line) is also shown. Specific results of S1 and S2 channels are also shown inside.
\label{FIG4}
}
\end{figure}

Fig.~\ref{FIG3} displays the evolutions of octupole deformations and pairing energies at $T$=0.9 and 1.5 MeV with
thermal fluctuations.
The resulted evolution times of different pathways are distributed widely.
At $T$=1.5 MeV, the fission now occurs with thermal fluctuations as an indispensible driving source.
The resulted scission deformations are widely distributed compared to that
of $T$=0.9 MeV, as a result of larger effects of thermal fluctuations at higher temperatures.
Pairing energies decrease at the beginning due to dissipations
and then induced dynamical pairing (not superfluid pairing) increases towards the scission due to thermal fluctuations,
exhibiting interesting competing roles of dissipation and fluctuation. The  induced pairing becomes prominent after long time evolutions.
It has also been shown that the reentrance of pairing
can happen in hot rotating nuclei~\cite{ddean}.

One of the key issues is the distributions of outcomes of TD-BCS calculations with thermal fluctuations.
Fig.\ref{FIG4} shows the fission pathways in the quadrupole-octupole deformation space.
At  $T$=0.75 MeV ($E^{*}$=10.6 MeV) and $T$=0.9 MeV ($E^{*}$=16.1 MeV), the fission yields are mainly distributed around two asymmetric channels.
For example, the average masses of heavy fragments at $T$=0.9 MeV  are around $A_H$=135.0 and 138.7 for S1 and S2 channels, respectively.
The associated average TKE are around 186.9 and 173.3 MeV respectively.
This is exactly the two standard asymmetric fission channels of  $^{240}$Pu in the Brosa model~\cite{brosa}.
The two channels of pathways are close in the deformation space while S2 corresponds to a larger deformation or a longer neck.
The onset of two asymmetric channels is mainly due to dynamical fluctuations while it would be difficult to identify them by models based on static PES.
It is understandable that the longer neck structure leads to smaller TKE and wider distributions. The longer S2 pathways also lead to
more dissipations and higher excitations of fragments (see the supplement~\cite{supp}), leading to the slope of the sawtooth structure of neutron multiplicities.
At $T$=1.5 MeV ($E^{*}$=53.2 MeV), the splitting of S1 and S2 is not clear any more. The distributions of scission deformations and masses
are much wider than that of 0.75 and 0.9 MeV.
 This demonstrated that the splitting of S1 and S2 disappears due to increasing fluctuations at high excitations.
 In Fig.\ref{FIG4}, S1 is dominated at $T$=0.75 MeV  and S2 is dominated at $T$=0.9 MeV.
Systematic analysis of experiments has also found  that S2 is dominated and the percentage of S1 channel decreases with increased energies~\cite{hujm}.

Finally, Table \ref{tab1} displays calculated fission observables.
 The complete results of all fluctuated pathways are given in the supplement~\cite{supp}.
In experiments, the averaged TKE of $^{239}$Pu($n,f$) is about 175 MeV and slightly decreases with increasing energies~\cite{tke}.
It is related to larger scission deformations and decreased S1 percentage at higher excitations as shown in Fig.\ref{FIG4}.
The experimental averaged mass of  heavy fragments  $A_H$ is about 140~\cite{brosa} rather than the magic number 132.
It is shown that TKE and $A_H$  from TD-BCS with temperatures and TDHF are about 186 MeV and 135.5, which are around S1 channel.
On the other hand,  TKE and $A_H$  from TD-BCS with initial pairing are about 175 MeV and 138.
We see that a considerable initial pairing is favorable for S2 channel.
Without thermal fluctuations, the resulted fission observables are close to S1 mode.
With thermal fluctuations increase, S2 mode gradually takes over S1 mode and finally averaged TKE and $A_H$ come back to experiments with considerable spreading widths.
We have demonstrated the essential role of thermal fluctuations in fission of compound nuclei when initial pairings vanish
and dissipations increase.
It is also of great interests to obtain excitation energies of fragments, which are relevant to neutron multiplicities.
In Table \ref{tab1}, heavy fragments have less excitation energies at low excitations but become close to that of light fragments at high excitations, which is
reasonable as the sawtooth structure would fade away at high excitations~\cite{2020}.
In conclusion, it is promising to develop a unifying framework for various energy-dependent fission observables with more pathways, a suitable effective temperature for fluctuations and also varying initial deformations~\cite{bulgac2,paul}.
Our work sheds a new light on the intuitive Brosa model for extrapolations and provides valuable clues towards a predictive microscopic fission theory.

\acknowledgments
We are grateful to discussions with F.R.Xu and W. Nazarewicz, and also discussions in the workshop on ``Future of Theory in Fission" held
at University of York in October 2019.
 This work was supported by  the
 National Key R$\&$D Program of China (Contract No. 2018YFA0404403),
  the National Natural Science Foundation of China under Grants No. 11975032, 11835001, 11790325, 11961141003.
  It was also supported by UK STFC under grant number ST/P005314/1.
We also acknowledge that computations in this work were performed in Tianhe-1A
located in Tianjin.


\begin{thebibliography}{99}


\bibitem{meitner}
L. Meitner, O. R. Frisch, Disintegration of Uranium by Neutrons: a New Type of Nuclear Reaction,
Nature 143, 239 (1939).


\bibitem{wp}
L.A. Bernstein, D. A. Brown, A. J. Koning, B.T. Rearden, C. E. Romano, A. A. Sonzogni, A. S. Voyles, and W. Younes,
Our Future Nuclear Data Needs, Ann. Rev. Nucl. Part. Sci. 69, 109(2019).

\bibitem{she}
J.H. Hamilton, S. Hofmann, and Y.T. Oganessian, Search for Superheavy Nuclei, Ann. Rev. Nucl. Part. Sci. 63, 383(2013).

\bibitem{pei2009}
J.C. Pei, W. Nazarewicz, J.A. Sheikh and A.K. Kerman, Fission Barriers of Compound Superheavy Nuclei,  Phys. Rev. Lett. 102, 192501(2009).

\bibitem{rp1}
M. Eichler, A. Arcones, A. Kelic, O. Korobkin, K. Langanke, T. Marketin, G. Martinez-Pinedo, I. V. Panov, T. Rauscher,
S. Rosswog, C. Winteler, N. T. Zinner, F.K. Thielemann, The Role of Fission in Neutron Star Mergers and its Impact on the r-Process Peaks,
Astrophys. J. 808, 30(2015).


\bibitem{rp2}
S. Goriely, The fundamental role of fission during r-process
nucleosynthesis in neutron star mergers, Eur. Phys. J. A 51, 22(2015).

\bibitem{witek}
J. Sadhukhan, S. A. Giuliani, Z. Matheson, and W. Nazarewicz, Efficient method for estimation of fission fragment yields of
r-process nuclei,
Phys. Rev. C 101, 065803(2020).


\bibitem{endf}
M.B. Chadwick, et al., ENDF/B-VII.1 Nuclear Data for Science and Technology: Cross Sections, Covariances, Fission Product Yields and Decay Data,
Nuclear Data Sheets 112,  2887 (2011).


\bibitem{yorkreview}
 M. Bender \textit{et al.}, Future of Nuclear Fission Theory, J. Phys. G 47, 113002(2020).


\bibitem{Schmidt}
K.H. Schmidt and B. Jurado, Review on the progress in nuclear
fssion--experimental methods
and theoretical descriptions, Rep. Prog. Phys. 81, 106301(2018).

\bibitem{schunck}
N.Schunck and L. M. Robledo,
Microscopic theory of nuclear fission: a review, Rep. Prog. Phys. 79 116301(2016).













\bibitem{bertsch}
M. Thoennessen and G. F. Bertsch, Threshold for dissipative fission,
Phys. Rev. Lett. 71, 4303 (1993).


\bibitem{dipole}
P. Paul and M. Thoennessen, Fission Time Scales from Giant Dipole Resonances, Annu. Rev. Nucl. Part. Sci. 44, 65(1994).


\bibitem{fluc-diss}
P.Fr\"{o}brich, I.I.Gontchar, N.D.Mavlitov, Langevin fluctuation-dissipation dynamics of hot nuclei: Prescission neutron multiplicities and fission probabilities,
Nucl. Phys. A 556, 281(1993).

\bibitem{zhuy}
Y. Zhu, and J. C. Pei, Thermal fission rates with temperature dependent fission barriers,
Phys. Rev. C 94, 024329 (2016).


\bibitem{tdgcm2}
J. Zhao, T. Nik\v{s}i\'{c}, D. Vretenar, and S.G. Zhou, Microscopic self-consistent description of induced fission dynamics: Finite-temperature effects,
Phys. Rev. C 99 014618 (2019).


\bibitem{2017}
D. E. Ward, B. G. Carlsson, T. D{\o}ssing, P. M?ller, J. Randrup, and S. {\AA}berg, Nuclear shape evolution based on microscopic level densities,
Phys. Rev. C 95, 024618(2017)


\bibitem{brosa}
U.Brosa, S.Grossmann, A. M\"{u}ller, Nuclear Scission, Phys. Rept. 197, 167(1990).

\bibitem{scamps}
G. Scamps, C. Simenel, Impact of pear-shaped fission fragments on mass-asymmetric fission in actinides, Nature 564, 382 (2018).

\bibitem{tdgcm}
D. Regnier, N. Dubray, N. Schunck, and M. Verriere, Fission fragment charge and mass distributions in
239Pu(n,f)
 in the adiabatic nuclear energy density functional theory,
Phys. Rev. C 93, 054611(2016).


\bibitem{langevin}
K. Sekimoto, Langevin Equation and Thermodynamics, Prog. Theo. Phys. Supp. 130, 17(1998).

\bibitem{langevin2}
J. Randrup and P. M\"{o}ller, Energy dependence of fssion-fragment mass distributions
from strongly damped shape evolution,  Phys. Rev. C 88, 064606(2013).


\bibitem{langevin3}
L.L. Liu, X.Z. Wu, Y.J. Chen, C.W. Shen, Z.X. Li, and Z.G. Ge, Study of fission dynamics with a three-dimensional Langevin approach,
Phys. Rev. C 99, 044614 (2019).



\bibitem{koonin}
S. E. Koonin and J. R. Nix, Microscopic calculation of nuclear dissipation,
Phys. Rev. C 13, 209(1976).

\bibitem{negele}
J. W. Negele, S. E. Koonin, P. Moller, J. R. Nix, and A. J. Sierk, Dynamics of induced fission,
Phys. Rev. C 17, 1098(1978).



\bibitem{tddft}
T. Nakatsukasa, K. Matsuyanagi, M.Matsuo, and K. Yabana, Time-dependent density-functional description of nuclear dynamics,
Rev. Mod. Phys. 88, 045004 (2016).

\bibitem{tdhf3}
C. Simenel, A.S. Umar, Heavy-ions collisions and fission dynamics with the time-dependent Hartree-Fock theory and its extensions,
 Prog. Part. Nucl. Phys. 103, 19 (2018).



\bibitem{bulgac2016}
A. Bulgac, P. Magierski, K. J. Roche, and I. Stetcu, Induced Fission of
240Pu
 within a Real-Time Microscopic Framework,
Phys. Rev. Lett. 116, 122504 (2016).



\bibitem{bulgac2}
A. Bulgac, S. Jin, K.J. Roche, N. Schunck, and I.Stetcu, Fission dynamics of
240Pu from saddle to scission and beyond,
Phys. Rev. C 100, 034615(2019).

\bibitem{258fm}
G. Scamps, C. Simenel, and D. Lacroix, Superfluid dynamics of 258Fm fission,
Phys. Rev. C 92, 011602(R) (2015).




\bibitem{lacroix2014}
D.  Lacroix, and S. Ayik, Stochastic quantum dynamics beyond mean field, Eur. Phys. J. A 50: 95(2014).


\bibitem{stdhf}
S. Ayik, A stochastic mean-field approach for nuclear dynamics, Phys. Lett. B 658,  174(2008).



\bibitem{bulgac3}
A. Bulgac, S. Jin, and I. Stetcu, Unitary evolution with fluctuations and dissipation,
Phys. Rev. C 100, 014615 (2019).


\bibitem{tdrpa}
R. Balian  and M. V\'{e}n\'{e}roni, Fluctuations in a time-dependent mean-field approach,
Phys. Lett. B 136, 301(1984).

\bibitem{176Yb}
K. Godbey, C. Simenel, and A. S. Umar, Microscopic predictions for the production of neutron-rich nuclei in the reaction
$^{176}$Yb+$^{176}$Yb,
Phys. Rev. C 101, 034602 (2020).


\bibitem{TDHFB}
S. Ebata, T. Nakatsukasa, T. Inakura, K. Yoshida, Y. Hashimoto, and K. Yabana,
Canonical-basis time-dependent Hartree-Fock-Bogoliubov theory and linear-response calculations,
Phys. Rev. C 82, 034306 (2010).

\bibitem{TDBCS}
G. Scamps, D. Lacroix, G. F. Bertsch, and K. Washiyama, Pairing dynamics in particle transport,
Phys. Rev. C 85, 034328 (2012).


\bibitem{FT-HFB}
A. L. Goodman, Finite-temperature HFB theory, Nucl. Phys. A 352, 30 (1981).


\bibitem{etdhf}
C.Y. Wong, H.H.K. Tang, Dynamics of nuclear fluid. V. Extended time-dependent Hartree-Fock approximation
illuminates the approach to thermal equilibrium, Phys. Rev. C 20, 1419(1979).

\bibitem{2016}
J. Sadhukhan, W. Nazarewicz, and N. Schunck, Microscopic modeling of mass and charge distributions in the spontaneous fission of
240Pu,
Phys. Rev. C 93, 011304(R)(2016).

\bibitem{sky3d}
J. A. Maruhn, P.-G. Reinhard, P. D. Stevenson, and A. S. Umar, The TDHF code Sky3D,
Comp. Phys. Comm. 185, 2195 (2014).

\bibitem{sky3d1}
B. Schuetrumpf, P.-G. Reinhard, P. D. Stevenson, A. S. Umar, and J. A. Maruhn, The TDHF code Sky3D version 1.1,
Comp. Phys. Comm. 229, 211 (2018).


\bibitem{skyax}
P.-G. Reinhard, B. Schuetrumpf, and J. A. Maruhn, The Axial Hatree-Fock + BCS Code SkyAx, Comp. Phys. Commun. (2020) doi:10.1016/j.cpc.2020.107603

\bibitem{marko}
M. Pancic, Y.Qiang, J.C. Pei, P. Stevenson, Shape Evolutions in Fission
Dynamics Within Time-Dependent
Hartree-Fock Approach, Front. Phys. 8, 351(2020).


\bibitem{skm}
J. Bartel, P. Quentin, M. Brack, C. Guet, and H. B. H{\aa}kansson, Towards a better parametrisation of Skyrme-like effective forces: A Critical study of the SkM force,
Nucl. Phys. A {\bf 386}, 79 (1982).


\bibitem{mix-pair}
{J. Dobaczewski, W. Nazarewicz, and M.V. Stoitsov, Nuclear ground-state properties from mean-field calculations,
Eur. Phys. J. A {\bf 15}, 21
  (2002)}.

\bibitem{supp}
See the Supplement Material for detailed methods and fission observables of individual events obtained from TD-BCS calculatoins with thermal fluctuations.


\bibitem{bender}
M. Bender, K. Rutz, P.-G. Reinhard, J. A. Maruhn, and W. Greiner, Potential energy surfaces of superheavy nuclei,
Phys. Rev. C 58, 2126 (1998).

\bibitem{timescale}
C. Simenel, K. Godbey, and A.S. Umar, Timescales of Quantum Equilibration, Dissipation and Fluctuation in Nuclear Collisions,
Phys. Rev. Lett. 124, 212504(2020).


\bibitem{feng}
P.H. Chen, Z.Q. Feng, J.Q. Li, H.F. Zhang, A statistical approach to describe highly excited heavy and
superheavy nuclei, Chin. Phys. C 40, 091002(2016).


\bibitem{khan}
E. Khan, N. Van Giai, N. Sandulescua, Pairing interactions and vanishing pairing correlations in hot nuclei,
Nucl. Phys. A 789, 94(2007).


\bibitem{paul}
P. Goddard, P. Stevenson, and A. Rios, Fission dynamics within time-dependent Hartree-Fock: Deformation-induced fission,
Phys. Rev. C 92, 054610 (2015).


\bibitem{ddean}
D. J. Dean, K. Langanke, H. Nam, and W. Nazarewicz, Pairing Reentrance Phenomenon in Heated Rotating Nuclei in the Shell-Model Monte Carlo Approach,
Phys. Rev. Lett. 105, 212504 (2010).


\bibitem{hujm}
U. Brosa, H.H. Knitter, T.S. Fan, J.M. Hu, S.L. Bao, Systematics of fission-channel probabilities, Phys. Rev. C 59, 767 (1999).







\bibitem{tke}
K. Meierbachtol, F. Tovesson, D. L. Duke, V. Geppert-Kleinrath, B. Manning, R. Meharchand, S. Mosby, and D. Shields,
Total kinetic energy release in
239Pu(n,f) post-neutron emission from 0.5 to 50 MeV incident neutron energy,
Phys. Rev. C 94, 034611(2016).


\bibitem{2020}
M.Albertsson, B.G.Carlsson, T.D{\o}ssing, P.M\"{o}ller, J.Randrup, S.{\AA}berg,
Excitation energy partition in fission,
Phys. Lett. B 803, 135276(2020).





\end{thebibliography}
\end{document}